\documentclass[referee]{cjaa}           

\usepackage{graphicx}                   
\input{epsf.sty}                        
\input{psfig.sty}                       

\begin{document}

   \title{High-energy Astrophysics and the Virtual Observatory
}


   \author{Paolo Padovani
      \inst{}\mailto{}
      }
   \offprints{Paolo Padovani}                   

   \institute{European Southern Observatory,
             Karl-Schwarzschild-Str. 2, D-85748 Garching bei
             M\"unchen, Germany\\
             \email{Paolo.Padovani@eso.org}
          }

   \date{Received~~2005 month day; accepted~~2005~~month day}

   \abstract{The Virtual Observatory (VO) will revolutionise the way
we do Astronomy by allowing easy access to all astronomical data and
by making the handling and analysis of datasets at various locations
across the globe much simpler and faster. I report here on the need
for the VO and its status in Europe, concentrating on the recently started
EURO-VO project, and then give two specific 
applications of VO tools to high-energy astrophysics.
}
   \authorrunning{}  
   \titlerunning{}  


   \maketitle
%
%
\section{Astronomy in the XXI century}       
\label{sect:intro}
Astronomy is facing the need for radical changes. When dealing with surveys
of up to $\sim 1,000$ sources, one could apply for telescope time and
obtain an optical spectrum for each one of them to identify the whole
sample. Nowadays, we have to deal with huge surveys (e.g., the Sloan
Digital Sky Survey [SDSS; \cite{{Abazajian04}}], the Two Micron All Sky
Survey [2MASS; \cite{Cutri03}], the Massive Compact Halo Object [MACHO;
e.g., \cite{Alcock04}] survey), reaching (and surpassing) the 100 million
objects. Even at, say, 3,000 spectra at night, which is only feasible with
the most efficient multi-object spectrographs and for relatively bright
sources, such surveys would require more than 100 years to be completely
identified, a time which is clearly much longer than the life span of the
average astronomer! But even taking a spectrum might not be enough to
classify an object. We are in fact reaching fainter and fainter sources,
routinely beyond the typical identification limits of the largest
telescopes available (approximately 25 magnitude for 2 - 4 hour exposures),
which makes ``classical'' identification problematic. These very large
surveys are also producing a huge amount of data: it would take almost  
three months to download at 1 Mbytes/s (an extremely good rate for most
astronomical institutions) the Data Release 4 (DR4; {\tt
http://www.sdss.org/dr4/}) SDSS images, more than a month for the
catalogues. The images would fill up $\sim$ 1,700 DVDs ($\sim$ 800 if using
dual-layer technology). And the final SDSS will be about 50 $\%$ as
large as the DR4. These data, once downloaded, need also to be analysed,
which requires tools which may not be available locally and, given the
complexity of astronomical data, are different for different energy
ranges. Moreover, the breathtaking capabilities and ultra-high efficiency
of new ground- and space-based observatories have led to a ``data
explosion'', with astronomers world-wide accumulating $\approx 1$ Terabyte
of data per night. For example, the European Southern Observatory
(ESO)/Space Telescope European Coordinating Facility (ST-ECF) archive is
predicted to increase its size by two orders of magnitude in the next seven
years or so, surpassing $\approx 1,000$ Terabytes. Finally, one would like to
be able to use all of these data, including multi-million-object
catalogues, by putting this huge amount of information together in a
coherent and relatively simple way, something which is impossible at
present.

All these hard, unescapable facts call for innovative solutions. For
example, the observing efficiency can be increased by a clever
pre-selection of the targets, which will require some ``data-mining'' to
characterise the sources' properties before hand, so that less time is
``wasted'' on sources which are not of the type under investigation. One
can expand this concept even further and provide a ``statistical''
identification of astronomical sources by using all the available,
multi-wavelength information without the need for a spectrum. The
data-download problem can be solved by doing the analysis where the data
reside. And finally, easy and clever access to all astronomical data
worldwide would certainly help in dealing with the data explosion and would
allow astronomers to take advantage of it in the best of ways.

\section{The Virtual Observatory}
The name of the solution is the Virtual Observatory (VO). The VO is an
innovative, evolving system, which will allow users to interrogate multiple
data centres in a seamless and transparent way, to utilise at best
astronomical data. Within the VO, data analysis tools and models,
appropriate to deal also with large data volumes, will be made more
accessible. New science will be enabled, by moving Astronomy beyond
``classical'' identification with the characterisation of the properties of
very faint sources by using all the available information. All this will
require good communication, that is the adoption of common standards and
protocols between data providers, tool users and developers. This is being
defined using new international standards for data access and mining
protocols under the auspices of the recently formed International Virtual
Observatory Alliance (IVOA: {\tt http://ivoa.net}), a global collaboration
of the world's astronomical communities.

One could think that the VO will only be useful to astronomers who deal
with colossal surveys, huge teams, and Terabytes of data. That is not the
case, for the following reason. The World Wide Web is equivalent to having
all the documents of the world inside one's computer, as they are all
reachable with a click of a mouse. Similarly, the VO will be like having
all the astronomical data of the world inside one's desktop. That will
clearly benefit not only professional astronomers but also anybody
interested in having a closer look at astronomical data. Consider the
following example: imagine one wants to find {\it all} data at {\it all}
wavelengths for a given high-energy source. One also needs to know which
ones are in raw or processed format, one wants to retrieve them and, if
raw, one wants also to have access to the tools to reduce them
on-the-fly. At present, this is extremely time consuming, if at all
possible, and would require, even to simply find out what is available, the
use a variety of search interfaces, all different from one another and
located at different sites. The VO will make it possible very easily.

\section{The Virtual Observatory in Europe}
The status of the VO in Europe is very good. In addition to seven current
national VO projects, the European funded collaborative Astrophysical
Virtual Observatory initiative (AVO: {\tt http://www.euro-vo.org}) had the
task of creating the foundations of a regional scale infrastructure by
conducting a research and demonstration programme on the VO scientific
requirements and necessary technologies. The AVO had been jointly funded by
the European Commission (under the Fifth Framework Programme [FP5]) with
six European organisations participating in a three year Phase-A work
programme (2001 -- 2004). The partners included ESO, the European Space
Agency (ESA), AstroGrid (funded by PPARC as part of the United Kingdom's
E-Science programme), the CNRS-supported Centre de Donn\'ees Astronomiques
de Strasbourg (CDS), the TERAPIX astronomical data centre at the Institut
d'Astrophisique in Paris, and the Jodrell Bank Observatory of the Victoria
University of Manchester. The AVO project is now formally concluded. Links
to various documents and to the software download page can be found at {\tt
http://www.euro-vo.org/twiki/bin/view/Avo/}.

\subsection{AVO's Main Achievements}\label{achieve}
AVO's main achievements can be thus summarised: 

\begin{enumerate}

\item {\it Science demonstrations}. The AVO project was driven by its
strategy of regular scientific demonstrations of VO technology. These were
held on an annual basis for its Science Working Group (SWG), established to
provide scientific advice to the project, in coordination with the
IVOA. Three very successful demonstrations were held in January 2003 (Jodrell Bank),
2004 (ESO, Garching), and 2005 (ESAC, Madrid).

\item {\it First VO paper}. The extragalactic case of the January 2004
demonstration was so successful that it turned into the first published
science result fully enabled via end-to-end use of VO tools and systems
(Padovani et al. 2004). The paper, which resulted in the discovery of $\sim
30$ high-power, supermassive black holes in the centres of apparently
normal looking galaxies, was also publicised by an ESA/ESO press release
(see {\tt
http://www.euro-vo.org/pub/articles/AVO1stSciencePressRelease.html}).

\item {\it VO tools}. For the purpose of the demonstrations progressively
more complex AVO demonstrators have been constructed. The current one is an
evolution of Aladin, developed at CDS, and has become a set of various
software components, provided by AVO and international partners, which
allows relatively easy access to remote data sets, manipulation of image
and catalogue data, and remote calculations in a fashion similar to remote
computing. The AVO prototype is a VO tool which can be used now for the
day-to-day work of astronomers. A Java application, it can be downloaded
from the AVO Web site at {\tt
http://www.euro-vo.org/twiki/bin/view/Avo/SwgDownload}.

\item {\it Science Reference Mission}. The Science Reference Mission is a
definition of the key scientific results that the full-fledged EURO-VO
should achieve when fully implemented. It consists of a number of science
cases, with related requirements, against which the success of the EURO-VO
will be measured. It was put together by the AVO Science Working Group.

\end{enumerate}

\subsection{The EURO-VO} 
The EURO-VO work program is the logical next step from AVO as a Phase-B
deployment of an operational VO in Europe. Building on the development
experience gained within the AVO Project, in coordination with the European
astronomical infrastructural networks OPTICON and RADIONET, and through
membership and support of the IVOA, EURO-VO will seek to obtain the
following objectives: 1) technology take-up and full VO compliant data and
resource provision by astronomical data centres in Europe; 2) support to
the scientific community to utilise the new VO infrastructure through
dissemination, workshops, project support, and VO facility-wide resources
and services; 3) building of an operational VO infrastructure in response
to new scientific challenges via development and refinement of VO
components, assessment of new technologies, design of new components and
their implementation. EURO-VO is open to all European astronomical data
centres. Initial partners include ESO, ESA, and six national nodes:
AstroGrid (UK), French-VO, GAVO (Germany), INAF (Italy), INTA (Spain), and
NOVA (The Netherlands). 

EURO-VO will seek to obtain its objectives by establishing three new
interlinked structures.

The EURO-VO Data Centre Alliance (DCA): An alliance of European data
centres who will populate the EURO-VO with data, provide the physical
storage and computational fabric and who will publish data, metadata and
services to the EURO-VO using VO technologies.

The EURO-VO Facility Centre (VOFC): An organisation that provides the
EURO-VO with a centralised registry for resources, standards and
certification mechanisms as well as community support for VO technology
take-up and dissemination and scientific program support using VO
technologies and resources.

The EURO-VO Technology Centre (VOTC): A distributed organisation that
coordinates a set of research and development projects on the advancement
of VO technology, systems and tools in response to scientific and community
requirements. 

The DCA will be a persistent alliance of data centre communities
represented at a national level. Through membership in the DCA, a nation's
community of data curators and data service providers will be represented
in a forum that will facilitate the take-up of VO standards, share best
practice for data providers, consolidate operational requirements for
VO-enabled tools and systems and enable the identification and promotion of
scientific requirements from programs of strategic national interest that
require VO technologies and services. Funds for the DCA will be requested
in an FP6 proposal to be submitted in September 2005.

The VOFC will provide a "public face" to the EURO-VO. Through outreach,
support of VO-enabled science projects in the community, workshops and
schools, the VOFC will represent a central support structure to facilitate
the broad take-up of VO tools by the community. The VOFC will also support
the EURO-VO Science Advisory Committee (SAC) to ensure appropriate and
effective scientific guidance from the community of leading researchers
outside the mainstream VO projects. The SAC will provide an up-to-date
stream of high-level science requirements to the EURO-VO. The VOFC will
further provide central services to the DCA for resource registry, metadata
standards and EURO-VO access. Funding for the VOFC has yet to be fully
defined but will come partially from ESO and ESA with activities ramping
up in 2006. 

The first VOFC activity was the organisation of a EURO-VO workshop at ESO
Headquarters in Garching from June 27 to July 1, 2005. The workshop was
explicitly designed for data centres and large projects to acquire the
knowledge and experience necessary to allow them to become "publishers" in
the VO. In tutorials and lectures, participants were instructed in the use
of VO analysis tools, libraries, and the existing web service
infrastructure to build VO compliant services. The workshop was aimed at
software engineers and designers building archive interfaces, writing
applications accessing remote data, or designing archive facilities and
data flows for future instruments and missions. More than 120 participants,
coming from 47 different institutions and 16 countries, attended the
workshop, with representatives from 11 out of 15 IVOA members. The tutorial
material is a collection of software which, although still not in a final
state, represents a unique and up-to-date snapshot of "state of the art" VO
technology. The workshop agenda and contributions are available at 
{\tt http://www.euro-vo.org/workshop2005}. 

The VOTC will consist of a series of coordinated technology research and
development projects conducted in a distributed manner across the member
organisations. The first project under the VOTC is the VO-TECH project,
funded through the EC FP6 Proposal and contributions from the Universities
of Edinburgh, Leicester, and Cambridge in the United Kingdom, ESO, CNRS
(France), and INAF (Italy). It is envisioned that additional projects will
be brought to the VOTC via other member organisations. The VOTC provides a
mechanism to coordinate and share technological developments, a channel for
DCA and VOFC requirements to be addressed and for technological
developments to be distributed to the community of data centres and
individual scientists in a coordinated and effective manner.

The EURO-VO project will be proactive in reaching out to European
astronomers. As a first step, the EURO-VO will be making regular
appearances at Joint European and National Astronomy Meetings (JENAM),
starting with the one in Liege in 2005.

\section{The VO and high-energy astrophysics}
I present now two applications of the usage of Virtual Observatory
tools to the field of high-energy astrophysics, namely: an example of
"data discovery" for a given astronomical source; and a specific
science case, the discovery of optically faint, obscured quasars in
the Great Observatories Origin Deep Survey (GOODS) fields.

\subsection{Data discovery in high-energy astrophysics: NGC 1068}
To give a (small) flavour of the possibilities offered by VO tools I have
used the AVO prototype (v. 2.002) (see Sect. \ref{achieve}) to look for
astronomical data for the prototype Seyfert 2 galaxy NGC 1068.

\begin{figure}
   \vspace{2mm}
   \begin{center}
 \hspace{3mm}\psfig{figure=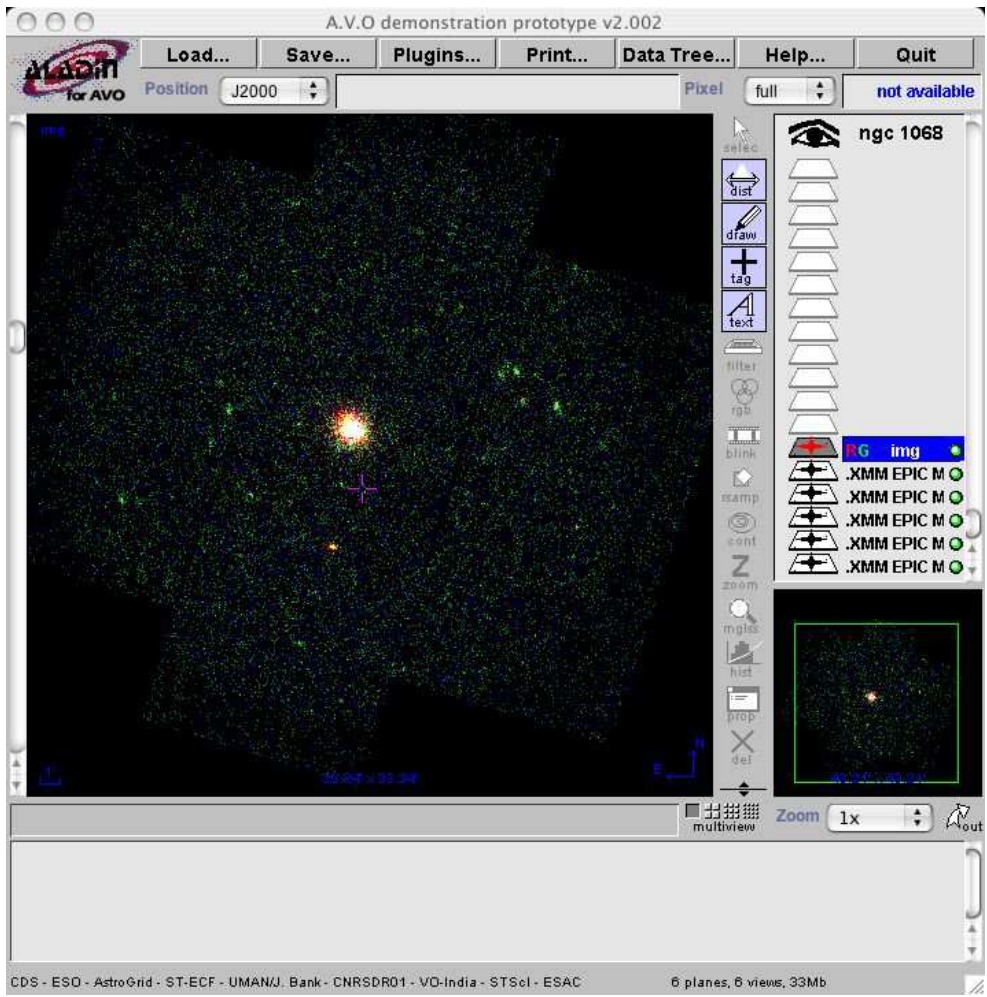,width=130mm,angle=0.0}  \parbox{180mm}{{\vspace{2mm} }}
   \caption{An X-ray colour composite view of NGC 1068 built from three XMM
   EPIC images in the $0.5 - 2.0$ (red), $2.0 - 4.5$ (green), and $4.5 -
   7.5$ keV (blue) band.}\label{rgb} \end{center}
\end{figure}

We start by looking for X-ray Multi Mirror (XMM) data for this source. We
do that by using the "VOdemo" button (which appears after clicking on
"Load") and then by selecting the "SIA server for XMM-Newton archive"
(where "SIA" stands for "Simple Image Access", an IVOA standard to access
images). After typing the name of our source we are presented with a list
of XMM images available directly from the XMM archive. We select five of
them in five different bands: $0.2 - 0.5$, $0.5 - 2.0$, $2.0 - 4.5$, $4.5 -
7.5$, and $7.5 - 12.0$ keV.  These can then be downloaded and displayed
individually or at the same time using the "multiview" feature of the
prototype. Three of them can then be combined using the "rgb" button to
create a colour composite image of the field around our source to put into
evidence, for example, the softest or hardest sources in the field
(Fig. \ref{rgb}).

\begin{figure}
   \vspace{2mm}
   \begin{center}
 \hspace{3mm}\psfig{figure=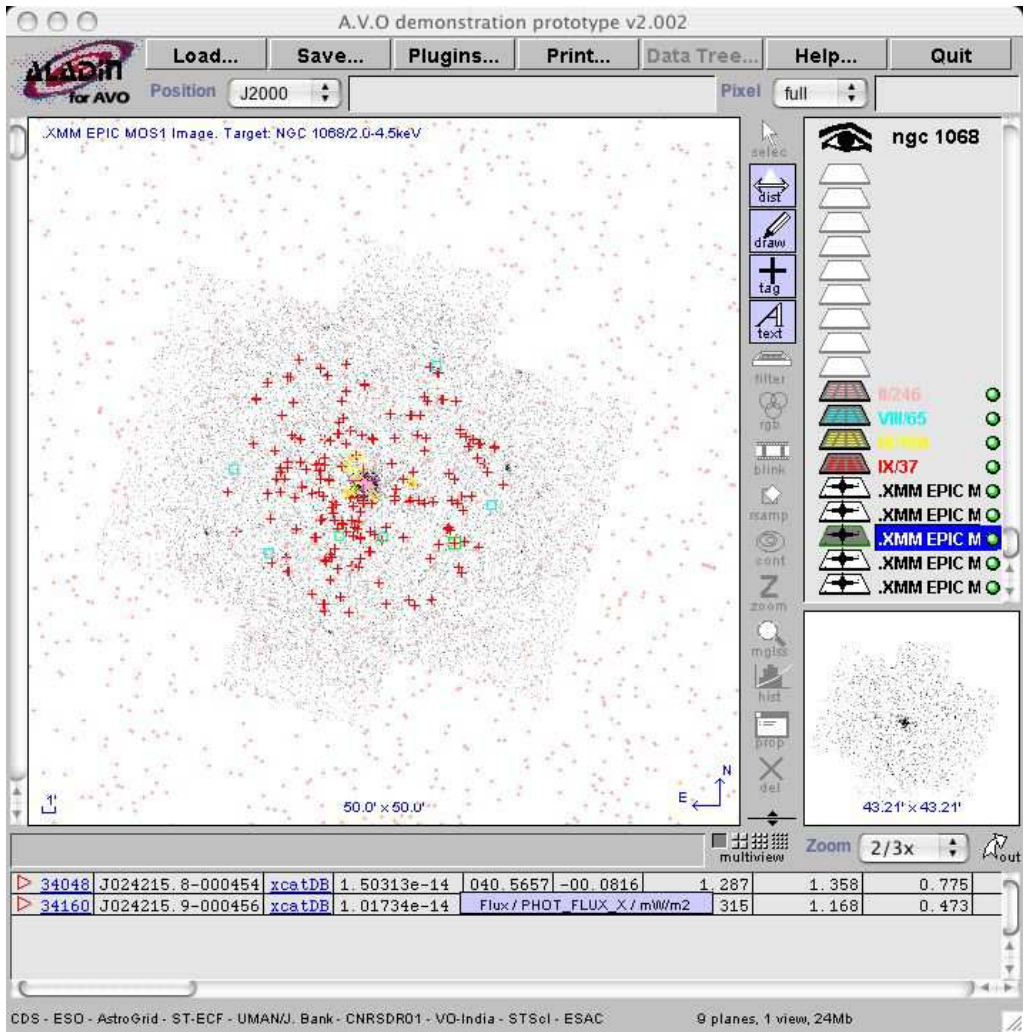,width=135mm,angle=0.0}  \parbox{180mm}{{\vspace{2mm} }}
   \caption{An XMM EPIC image of NGC 1068 with entries from the 2MASS,
   NVSS, RASS, and 1XMM catalogues overlaid.}\label{catalogs} \end{center}
\end{figure}

One can then look for various astronomical data for NGC 1068 and
neighbouring sources.  The AVO prototype has a direct link to VizieR ({\tt
http://vizier.u-strasbg.fr/}), a service which provides access to the most
complete library of published astronomical catalogues and data tables
available on-line. We can then search for all VizieR catalogues by
wavelength, mission, keyword, or simply author name. For example, we might
want to look for 2MASS, NRAO-VLA Sky Survey (NVSS), {\it ROSAT} All-Sky
Survey (RASS), and 1XMM sources in a region around NGC 1068
(Fig. \ref{catalogs}). The corresponding catalogues get loaded into the
prototype and the sources are automatically overlaid on the selected image
with different symbols and colours corresponding to different
catalogues. By selecting some sources with the cursor these get highlighted
and the catalogue parameters appear in the window below the image. At this
point one might want to plot catalogue entries, which can be done by using
VOPlot, the graphical plug-in of the prototype, or create new columns by
doing numerical manipulations on the existing ones. Or one can also
cross-correlate the catalogues to find out, for example, which of the 2MASS
sources have, or do not have, an XMM counterpart by using the catalogue
cross match tool.

Finally, one might want to look for available spectra of NGC 1068 in
various bands. This can be accomplished using VOSpec, a standalone tool
which is also one of the plug-ins of the prototype. At present, VOSpec
looks for spectra for a selected astronomical object (or around a position)
in various archives which follow the Simple Spectrum Access (SSA) protocol,
an IVOA standard for spectra.  These include: the Infrared Space
Observatory (ISO), the Hubble Space Telescope (HST) archive at the Space
Telescope European Coordinating Facility (ST-ECF) (for re-calibrated Faint
Object Spectrograph [FOS] spectra), the HyperLeda archive, the IUE Newly
Extracted Spectra (INES) archive, the Far Ultraviolet Spectroscopic
Explorer (FUSE) archive in Paris, the GIRAFFE archive, and the SDSS
archive. Once the spectra are located VOSpec presents the user with a list
of them.  The selected files are then retrieved and plotted to build a
Spectral Energy Distribution (SED).  VOSpec manages to superimpose spectra
coming from different instruments and energy bands by converting all
spectra to the same units, using the information provided by the archives
as part of the SSA protocol.

\begin{figure}
   \vspace{2mm}
   \begin{center}
 \hspace{3mm}\psfig{figure=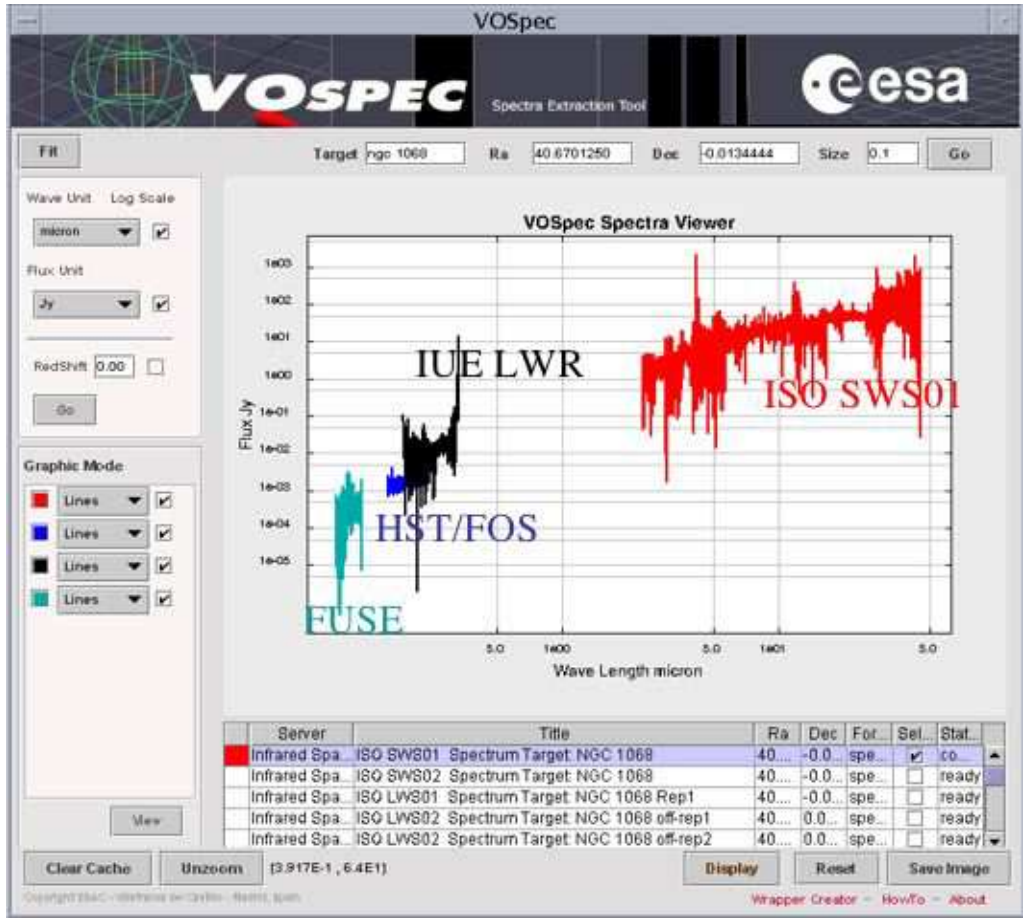,width=135mm,angle=0.0}  \parbox{180mm}{{\vspace{2mm} }}
   \caption{An example of a composite spectrum of NGC 1068 built using the VOSpec tool and,  going from shorter to longer wavelengths (left to right), FUSE, HST/FOS, IUE Long Wavelength Redundant (LWR), and ISO Short Wavelength Spectrometer (SWS) data. }\label{vospec}
   \end{center}
\end{figure}

Figure \ref{vospec} gives an example of a composite spectrum of NGC 1068
built using, going from shorter to longer wavelengths (left to right),
FUSE, HST/FOS, IUE Long Wavelength Redundant (LWR), and ISO Short
Wavelength Spectrometer (SWS) data. Obviously astronomers need to be
cautious when merging spectra coming from different instrument and taken
with different apertures but VOSpec gives a very nice overview of the SED
of astronomical objects. This tool will obviously become even more useful
when more archives adopt the SSA protocol.

\subsection{Discovering optically faint, obscured quasars with VO tools}
The AVO held its second demonstration, 'AVO 1st Science', on January 27 -
28, 2004 at ESO. The demonstration was truly multi-wavelength, using
heterogeneous and complex data covering the whole electromagnetic
spectrum. These included: MERLIN, VLA (radio), ISO [spectra and images] and
2MASS (infrared), USNO, ESO 2.2m/WFI and VLT/FORS [spectra], and HST/ACS
(optical), XMM and Chandra (X-ray) data and catalogues. Two cases were
dealt with: an extragalactic case on obscured quasars, centred around the
the two GOODS fields (Giavalisco et al. 2004), namely the Hubble Deep
Field-North (HDF-N) and the Chandra Deep Field-South (CDF-S), the most
data-rich, deep survey areas on the sky; and a Galactic scenario on the
classification of young stellar objects.

The extragalactic case was so successful that it turned into the first
published science result fully enabled via end-to-end use of VO tools and
systems, the discovery of $\sim 30$ high-power, supermassive black holes in
the centres of apparently normal looking galaxies (Padovani et al. 2004). 

Black holes lurk at the centres of active galaxies (AGN) surrounded by dust
which is thought to be, on theoretical and observational grounds (see,
e.g., Urry \& Padovani 1995; Jaffe et al. 2004), distributed in a flattened
configuration, torus-like. When we can look down the axis of the dust torus
and have a clear view of the black hole and its surroundings these objects
are called ``type 1'' AGN, and display the broad lines and strong UV
emission typical of quasars. ``Type 2'' AGN, on the other hand, lie with
the dust torus edge-on as viewed from Earth so our view of the black hole
is totally blocked by the dust over a range of wavelengths from the
near-infrared to soft X-rays.

While many dust-obscured low-power black holes, the Seyfert 2s, have been
identified, until recently few of their high-power counterparts were
known. This was due to a simple selection effect: when the source is a
low-power one and therefore, on average, closer to the observer, one can
very often detect some features related to narrow emission lines on top of
the emission from the host galaxy, which qualify it as a type 2 AGN. But
when the source is a high-power one, a so-called QSO 2, and therefore, on
average, further away from us, the source looks like a normal galaxy.

Our approach was to look for sources where nuclear emission was coming out
in the hard X-ray band, with evidence of absorption in the soft band, a
signature of an obscured AGN, and the optical flux was very faint, a sign
of absorption. One key feature was the use of a correlation discovered by
Fiore et al. (2003) between the X-ray-to-optical ratio and the X-ray power,
which allowed us to select QSO 2s even when the objects were so faint that
no spectrum, and therefore no redshift, was available.

We used a large amount of data: deep X-ray (Chandra) and optical (HST/ACS)
catalogues, and identifications, redshifts, and spectra for previously
identified sources in the CDF-S and HDF-N based on VLT and Keck data.
Using the AVO prototype made it much easier to classify the sources we were
interested in and to identify the previously known ones, as we could easily
integrate all available information from images, spectra, and catalogues at
once. One interesting feature was the prototype catalogue cross-matching
service, which can access all entries in Vizier, at CDS, and allowed us to
cross-correlate a variety of catalogues very efficiently.

Out of the 546 X-ray sources in the GOODS fields we selected 68 type 2 AGN
candidates, 31 of which qualify as QSO 2 (estimated X-ray power $> 10^{44}$
erg/s). Our work brings to 40 the number of QSO 2 in the GOODS fields, 
an improvement of a factor $\sim 4$ when compared to the only nine
such sources previously known. These sources are very faint ($\langle R
\sim 27\rangle$) and therefore spectroscopical identification is not
possible, for the large majority of objects, even with the largest
telescopes currently available. By using VO methods we are sampling a
region of redshift - power space so far unreachable with classical methods.
For the first time, we can also assess how many QSO 2 there are down to
relatively faint X-ray fluxes. We find a surface density $> 330$ deg$^{-2}$
for $f(0.5 - 8 keV) \ge 10^{-15}$ erg cm$^{-2}$ s$^{-1}$, higher than some
previous estimates.

The identification of a population of high-power obscured black holes
and the active galaxies in which they live has been a key goal for
astronomers and will lead to greater understanding and a refinement of
the cosmological models describing our Universe. 

This result is proof that VO tools have evolved beyond the
demonstration level to become respectable research tools, as the VO is
already enabling astronomers to reach into new areas of parameter space
with relatively little effort. 


\section{Conclusions}\label{sec:concl}

The main conclusions are as follows:

\begin{enumerate}

\item We need to change the way we do Astronomy if we want to take
advantage of the huge amount of data we are being flooded with. The way to
do that is through the Virtual Observatory.

\item The Virtual Observatory will make the handling and analysis of
astronomical data and tools located around the world much easier, enabling
also new science.

\item Everybody will benefit, including high-energy astrophysics researchers!

\item Virtual Observatory tools are available now to facilitate
astronomical research and, as I have shown, can also be applied to high-energy astrophysics.

\end{enumerate}

Visit {\tt http://www.euro-vo.org/twiki/\-bin/\-view/Avo/SwgDownload}
to download the AVO prototype. I encourage astronomers to download the
prototype, test it, and use it for their own research. For any
problems with the installation and any requests, questions, feedback,
and comments you might have please contact the AVO team at
twiki@euro-vo.org. (Please note that this is still a prototype:
although some components are pretty robust some others are not.)

\begin{acknowledgements}
I thank the Astrophysical Virtual Observatory team for their superb work. I
have made extensive use of the CDS VizieR catalogue tool, SIMBAD and the
Aladin sky atlas service, and of the VOSpec tool developed at the European
Space Astronomy Centre (ESAC). The Astrophysical Virtual Observatory was
selected for funding by the Fifth Framework Programme of the European
Community for research, technological development and demonstration
activities, under contract HPRI-CT-2001-50030. The EURO-VO VO-TECH project
was selected for funding by the Sixth Framework Programme of the European
Community.
\end{acknowledgements}

\label{lastpage}

\end{document}